\documentclass[prl,twocolumn,aps]{revtex4}
\usepackage{verbatim}
\usepackage{graphicx}
\normalfont
\begin{document}
%\large

\title{Full coherent control of nuclear spins in an optically pumped single quantum dot}

%\wideabs{

\author{M. N. Makhonin$^1$, K. V. Kavokin$^{2}$, P. Senellart$^3$, A. Lema\^itre$^3$, A. J. Ramsay$^1$, M. S. Skolnick$^1$, A. I. Tartakovskii$^1$}

\address{$^{1}$ Department of Physics and Astronomy, University of Sheffield, S3 7RH,UK \\ $^{2}$ A. F. Ioffe Physico-Technical Institute, 194021, St. Petersburg, Russia,\\ $^3$ Laboratoire de Photonique et de Nanostructures, Route de Nozay, 91460 Marcoussis, France}
\date{\today}

%\begin{abstract}
%\end{abstract}

\maketitle

%}

{\bf Highly polarized nuclear spins within  a semiconductor quantum dot (QD) induce effective magnetic (Overhauser) fields of up to several Tesla acting on the electron spin \cite{Eble,Tartakovskii,Chekhovich1,Xu,Latta,Nikolaenko,Gammon1,Makhonin,Chekhovich2,Foletti,Kloeffel,KKM}, or up to a few hundred mT for the hole spin \cite{Chekhovich3,Fallahi}. Recently this has been recognized as a resource for intrinsic control of QD-based spin quantum bits. However, only static long-lived Overhauser fields could be used \cite{Foletti,Kloeffel}. Here we demonstrate fast redirection on the microsecond time-scale of Overhauser fields of the order of 0.5 T experienced by a single electron spin in an optically pumped GaAs quantum dot. This has been achieved using full coherent control of an ensemble of $10^3-10^4$ optically polarized nuclear spins by sequences of short radio-frequency (rf) pulses. These results open the way to a new class of experiments using rf techniques to achieve highly-correlated nuclear spins in quantum dots, such as adiabatic demagnetization in the rotating frame \cite{Slichter} leading to sub-$\mu$K nuclear spin temperatures, rapid adiabatic passage \cite{Slichter}, and spin squeezing \cite{Rudner}.}

\begin{figure}
\centering
\includegraphics[width=8cm]{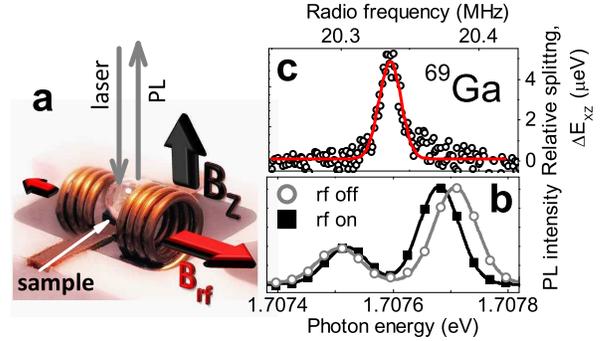}
\caption{Experimental techniques used for optically detected nuclear magnetic resonance (ODNMR) in a single dot. (a) The ODNMR experiment, showing the sample containing strain-free interface GaAs/AlGaAs dots. Optical excitation and photoluminescence (PL) collection is parallel to the external magnetic field $B_{z}$ along the sample growth axis $z$, whereas the radio-frequency field $B_{rf}$ is normal to $z$. PL is collected by a solid immersion lens placed on top of the sample. (b) PL spectra of the neutral exciton in a GaAs dot under circularly polarized excitation leading to non-zero Overhauser field $B_N$. The two peaks are the exciton Zeeman doublet, with the splitting dependent on both $B_N$ and external field $B_z$ ($B_z=2$T here). The spectrum plotted with black symbols is measured under simultaneous rf excitation, for which $B_N$ is partially suppressed. Grey symbols show the case when a larger value of $B_N$ is induced by the optical pumping only. (c) An ODNMR spectrum measured for $^{69}$Ga at $B_{z}=1.99$T (symbols) by performing measurements of the exciton Zeeman splitting $E$ as in (b) for a range of rf frequencies. The spectrum is fitted with a Gaussian (line) with a full width of 16 kHz.}
\label{fig1}
\end{figure}

A single electron spin confined in a semiconductor quantum dot (QDs) interacts with $10^4-10^6$ nuclear spins. As a result, the effective nuclear (Overhauser) field experienced by the electron spin influences its life-time \cite{Erlingsson}, coherence \cite{Khaetskii,Petta,Koppens}, and precession frequency   \cite{Eble,Tartakovskii,Chekhovich1,Xu,Latta,Nikolaenko,Gammon1,Makhonin,Chekhovich2,Foletti,Kloeffel,Petta,Koppens}. Recently,  quasi-static Overhauser fields have been used as a tool for the control of the double-dot two-electron qubit \cite{Foletti}, and slow tuning of the polarization and wavelength of the emitted single photons in an optically pumped dot \cite{Kloeffel}. However, manipulation of the Overhauser fields in quantum dots used indirect methods based on the hyperfine interaction with confined electrons. As a result the nuclear spin alignment parallel to the direction of the average electron spin occurs on time-scales of  a few seconds in non-zero fields \cite{Foletti,Latta,Chekhovich1,Nikolaenko,Vink}.

Here we report full coherent control of  nuclear spins in a single GaAs/AlGaAs  quantum dot in a magnetic field of a few Tesla. This enables rotation of large Overhauser fields about any axis on the micro-second scale. Such control is achieved by using a sequence of two phase-locked rf pulses to induce coherent rotations of a group of  $^{69}$Ga nuclear spins optically pumped to a high polarization degree.
A fast 15 $\mu$s $\pi$-rotation of the Overhauser field of $\approx$ 500 mT has been achieved on the time-scale of long-lived coherence up to 80 $\mu$s of an electron spin qubit \cite{Barthel}. The result of the spin rotations is measured from the nuclear spin polarization along the external magnetic field direction by detecting the Overhauser shifts in photoluminescence (PL) of the electron-hole pair in the dot (see Fig.1 and Methods). Analysis of Rabi oscillations and spin-echo data shows that the dynamic response of nuclear spins under pulsed rf excitation is sensitive to the inhomogeneous distribution of resonant frequencies in the spin ensemble. Prior to the present work, only quasi-continuous-wave nuclear magnetic resonance (NMR) experiments have been reported on single dots, allowing access to single electron properties \cite{Gammon1,Makhonin}. Coherent manipulation of nuclear spins using rf techniques has been demonstrated, but only for macroscopic systems containing in excess of $10^9$ nuclei and many weakly localized electrons \cite{Yusa,Machida,Sanada}.

A necessary first step towards full coherent control is rotation of nuclear spins using a single rf pulse. We use modulated techniques, as shown in Fig.2a (further experimental and sample details are given in Fig.1 caption and Methods). Rabi oscillations in Fig.2b are observed  by varying the duration $\tau_{p}$ of the rf pulse in resonance with $^{69}$Ga nuclei and measuring the change in the exciton Zeeman splitting $\Delta E$ using a delayed optical pulse. $\Delta E$ is directly proportional to the change of the $z$-projection of the Overhauser field $B_{Nz}$, which in its turn follows the change of the nuclear spin polarization along $z$ axis, $S_{Nz}$: $B_{Nz}\propto S_{Nz}$. The transverse component of $B_N$ in the XY plane decays on a time scale shorter than the delay between the rf and read-out optical pulse, so that only $B_{Nz}$ contributes to the change in the exciton Zeeman splitting $\Delta E$. Dephasing leading to the decay of the transverse component of $B_N$ arises from the inhomogeneous distribution of the resonance frequencies in the spin ensemble, and also contributes to the decay of the amplitude of the Rabi oscillations with time observed in Fig.2b. This will be discussed in detail below. Fig.2b shows Rabi oscillations measured for three powers of the rf pump, producing rf fields increasing from the bottom curve to the middle and top by a factor of 2 and 3, respectively. The absolute magnitude of the rf field $B_{rf}$ can be deduced from the period of oscillations,  $T=2\pi/\gamma B_{rf}$ ($\gamma$ - gyromagnetic ratio of the nuclear spin \cite{Slichter}). Thus, the curves with periods of $\approx 160 \mu$s, $\approx 80 \mu$s, and $\approx 53 \mu$s correspond to $B_{rf}$ of 0.6, 1.2 and 1.8 mT, respectively. The fastest $\pi$-pulse observed in our experiments was 15 $\mu$s, limited by the magnitude of the rf field which could be applied. In that measurement, a total $\Delta E=21\mu$eV was detected, corresponding to a change in the nuclear magnetic field felt by the electron of approximately 900 mT (for $|g_e|=0.4$). Fig.2b also shows that the amplitude of the Rabi oscillations decays with time, reflecting dephasing in the spin ensemble (to be discussed below).

\begin{figure}
\centering
\includegraphics[width=8cm]{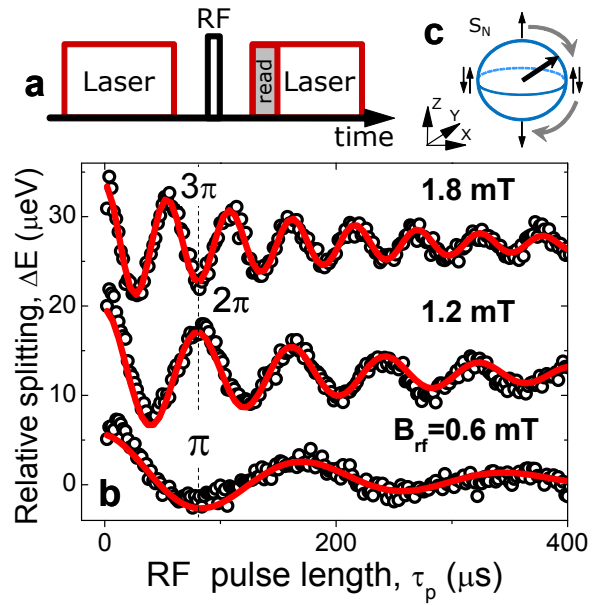}
\caption{Rabi oscillations of nuclear spins in a single GaAs dot at $B_z=3.55$T excited with a single rf pulse in resonance with $^{69}$Ga. (a) Diagram of the experimental cycle used for pulsed ODNMR experiments. 5 s laser pulses (red) are interrupted by 5 ms ''dark'' gaps when the rf pulse(s) of 1 to 1000 $\mu$s are applied (black). PL read-out used to measure changes of  the $z$-component of the total nuclear spin is carried out in the first 25 ms of the laser pulse after the rf excitation. (b) Rabi oscillations of nuclear spins observed as the variation of the exciton Zeeman splitting in the dot $E$ (denoted $\Delta E$) as a function of the rf pulse duration $\tau_{p}$ at $f_0=$36.3 MHz. Symbols show data measured for different rf powers corresponding to $B_{rf}$=0.6, 1.2 and 1.8 mT. The curves are displaced vertically for clarity. Lines show fitting using Eq.\ref{torrey} in the text, which takes into account both the intrinsic coherence time $T_2$ and the inhomogeneous broadening of the nuclear spin ensemble described by $T_2^*$. (c) shows a schematic diagram of a Bloch sphere and rotation of the nuclear spin around the rf field $B_{rf}$ directed along $Oy$.}
\label{fig2}
\end{figure}

\begin{figure}
\centering
\includegraphics[width=8.5cm]{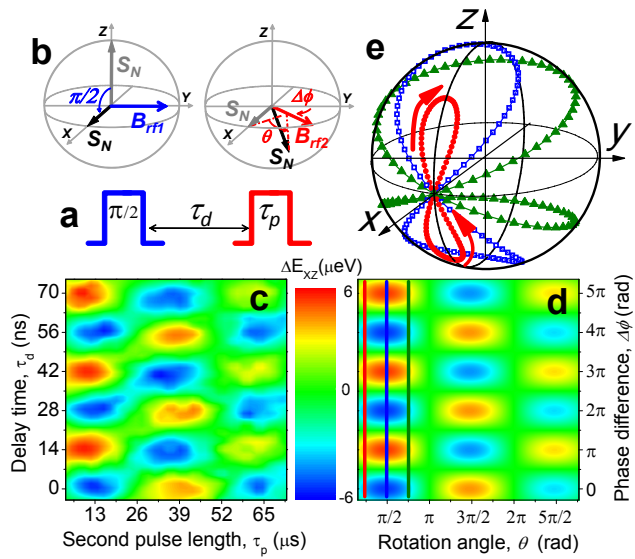}
\caption{ Full coherent control of nuclear spins in a dot. We use two phase-locked rf pulses: a $\pi/2$-pulse followed after a delay $\tau_d$  by a pulse with duration $\tau_p$ [see (a)]. (b) depicts rotations of the total nuclear spin $\mathbf{S_N}$ on the Bloch sphere due to excitation with: (i) a $\pi/2$-pulse with $\mathbf{B_{rf1}}$ along $Oy$ (left); (ii) a second phase-shifted pulse with $\mathbf{B_{rf2}}$ aligned at an angle $\Delta\phi$ with respect to $\mathbf{B_{rf1}}$ in XY plane (right). (c) presents the experimental demonstration of the full control of nuclear spins in a single quantum dot for $B_z=3.55$ T. The periodic variation of $\Delta E$ (or $z$-projection $S_{Nz}$ of $\mathbf{S_N}$) as a function of $\tau_p$ corresponds to spin rotation about a horizontal axis, the direction of which is defined by $\tau_d$, thus controlling rotation about $Oz$. (d) Simulations of the oscillations in (c) with temporal parameters $\tau_d$ and $\tau_p$ converted respectively to the phase-shift between the pulses $\Delta\phi$ and the angle $\theta$ of $\mathbf{S_N}$ rotation around $\mathbf{B_{rf2}}$. Decay of the Rabi oscillations as described by Eq.\ref{torrey} is taken into account. (e) Calculated trajectories of the end of the vector $\mathbf{S_N}$ on a Bloch sphere as a result of the two-pulse rotations. The trajectories are calculated  as a function of the phase between the pulses $\Delta\phi$ (i.e. $\tau_d$) for three different lengths of the second pulse, corresponding to spin rotation by $\theta=\gamma B_{rf2}\tau_p$ of $\pi/4$ (red), $\pi/2$ (blue) and $3\pi/4$ (green) around the rf field of the second pulse. This motion of $\mathbf{S_N}$ corresponds to the oscillations in $B_{Nz}$ along the three vertical lines in (d). Arrows next to the curve $\theta=\pi/4$ show the direction in which $\mathbf{S_N}$ moves as $\tau_d$ is increased. A similar ''8''-shape evolution is observed for the pulses with different durations.}
\label{fig3}
\end{figure}

Coherent manipulation of nuclear spins using two rf pulses enables fast realignment of the nuclear magnetic field in any given direction. We use a sequence of two rf pulses tuned into resonance with the $^{69}$Ga transition: a $\pi/2$ pulse followed after a delay $\tau_d$ by a pulse with a duration $\tau_{p}$ (see Fig.3a). Redirection of the Overhauser field is carried out by tuning of  $\tau_d$ and $\tau_{p}$, as explained below. The schematics in Fig.3b show a Bloch sphere in a frame rotating with the Larmor frequency $2\pi f_0=\gamma B_{z}$ around the external field, and the spin rotations produced by each pulse. The $\pi/2$ pulse with $\mathbf{B_{rf1}}$ along $Oy$ rotates the total nuclear spin $\mathbf{S_N}$ along $Ox$. The magnetic field of the second pulse, $\mathbf{B_{rf2}}$, is directed at an angle $\Delta\phi=2\pi f_0\tau_d$ with respect to that of the first pulse. Only the nuclear spin component normal to the direction of $\mathbf{B_{rf2}}$ is affected by this pulse: it rotates around this direction by an angle $\theta=\gamma B_{rf2}\tau_p$. As the $\tau_{p}$ is increased at a fixed $\Delta\phi$, periodic rotations of $\mathbf{S_N}$ around $\mathbf{B_{rf2}}$ will be observed, similar to the Rabi flops in Fig.2. These rotations will cause periodic changes of the $z$-projection of the Overhauser field $B_{Nz}$, measured in experiment. In their turn, the amplitude of these changes will be proportional to the magnitude of the spin projection normal to $\mathbf{B_{rf2}}$, $S_Ncos(\Delta\phi)$, dependent on the relative orientation of the magnetic fields of the two rf pulses. Therefore as the phase between the two rf pulses $\Delta\phi$ (or $\tau_{d}$) is varied at a fixed $\tau_{p}$, the amplitude of $B_{Nz}$ will also vary periodically exhibiting Ramsey fringes \cite{Ramsey}.

Such behavior is directly observed in experiment, where a pattern of periodic oscillations of the exciton Zeeman splitting $\Delta E\propto B_{Nz}$ is observed. The color-plot in Fig.3c shows the dependence of $\Delta E$ on the delay between the two pulses, $\tau_d$ defining the phase shift $\Delta\phi$, and the duration of the second pulse, $\tau_p$, defining the angle $\theta$ of the $\mathbf{S_N}$ rotation around $\mathbf{B_{rf2}}$. As predicted above, two types of oscillations are observed: (i) Rabi oscillations about a horizontal axis as a function of $\tau_p$ for a fixed phase difference $\Delta\phi$ (i.e. $\tau_d$) and (ii) Ramsey fringes describing periodic rotations around the vertical axis as a function of $\tau_d$ (or $\Delta\phi$) for a fixed duration of the second pulse. The experimental dependence is well reproduced by simulations (see Fig.3d), where we also take into account the decay of Rabi oscillations due to spin dephasing as further detailed below and described by Eq.\ref{torrey}. The motion of the total nuclear spin on the Bloch sphere corresponding to the periodic pattern observed in Fig.3c and 3d is further illustrated in Fig.3e, which also shows that by varying both $\tau_p$ and $\tau_{d}$, any point on the Bloch sphere can be accessed, constituting the full coherent control of nuclear spins.

\begin{figure}
\centering
\includegraphics[width=8cm]{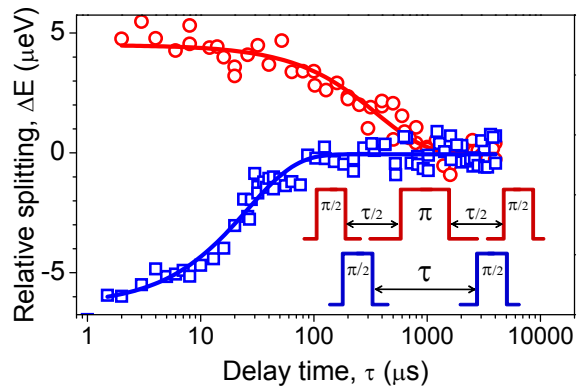}
\caption{Measurements of the intrinsic ($T_2$) and effective ($T_2^*$) spin coherence times in a single GaAs dot for $B_z=3.55$T. Diagrams show the sequences of pulses used for measurements of $T_2$ (spin-echo method shown schematically in red) and $T_2^*$  (excitation by two pulses, blue). All pulses in these sequences have the same phase. Circles (squares) show the dependence $E(\tau)$ for the three- (two-) pulse sequence. Fitting of the data with a mono-exponential decay (rise) yields $T_2=360\pm 40 \mu$s ($T_2^*=25\pm 2 \mu$s).}
\label{fig4}
\end{figure}

The efficiency of the pulse-sequence in controlling the Overhauser field, and also the dynamics of $B_N$ following the rotation are influenced by dephasing in the nuclear spin ensemble. There are two major contributions in our case: (i) fast dephasing due to the inhomogeneity of the nuclear spin ensemble and (ii) slow decoherence of individual spins due to dipole-dipole interaction between nuclear spins. The characteristic times of the processes (i) $T_2^*$ and (ii) $T_2$ are determined using the pulse sequences shown in Fig.4, where all pulses have the same phase. In the sequence $\pi/2-\tau-\pi/2$, nuclear spin polarization along $Oz$ decays with $T_2^*=25\pm 2 \mu$s as a function of $\tau$ due to the finite distribution of the spin precession frequencies in the ensemble (blue in Fig.4) \cite{Slichter}. This is in good agreement with the linewidth of $\approx 16$kHz observed in the  cw experiment (Fig.1c): $\Delta f=1/\pi T_2^*\approx 13$ kHz. In the spin-echo sequence $\pi/2-\tau/2-\pi-\tau/2-\pi/2$ \cite{Slichter}, the focusing $\pi$-pulse cancels this effect, and the intrinsic coherence time $T_2=360\pm 40 \mu$s is measured (red in Fig.4). The transverse component of the Overhauser field redirected by the two pulse-sequence in Fig.3 will decay with characteristic time $T_2^*$; however, the decay can be prolonged by applying the spin-echo method.

In the simplest case of negligible inhomogeneity in the spin ensemble the decay of Rabi oscillations should occur with a characteristic time $T_{Rabi}=2T_2\approx700 \mu$s \cite{Slichter,Abragam} (longitudinal spin decay time $T_1\approx 100$ s $\gg T_2$ in GaAs dots\cite{Nikolaenko}). However, the fitting of the Rabi oscillations in Fig.2 with a single exponential decay in the form $\Delta E=cos(\gamma B_{rf}\tau_p)exp(-\tau_p/T_{Rabi})$ yields a characteristic decay time $T_{Rabi}\approx 200 \mu$s. This discrepancy can be explained if the inhomogeneous width of the spin resonance is taken into account, which neglecting the complications due to the four states of the spin 3/2 Ga nucleus, can be described in terms of an ensemble of identical spins placed in a non-uniform magnetic field. The response of such a spin ensemble to the driving rf field has been described by Torrey in Ref.\cite{Torrey}, whose analytical expression we use to model the nuclear spins in the dot:  
\begin{equation}
\Delta E\propto \frac{exp(-\tau_p/T_2)}{[1+(\tau_p/T_H)^2]^{1/4}} cos(\gamma B_{rf}\tau_p)
\label{torrey}
\end{equation}
where $T_H=1.4 \gamma B_{rf} T_2^{*2}$. We fit all curves in Fig.2 using Eq.\ref{torrey} with adjustable parameters $B_{rf}$, $T_2^*$ and $T_2$. The best fit is obtained for $T_2^*=16 \pm 1 \mu$s and $T_2 = 310 \pm 30 \mu$s in good agreement with the independent measurements in Fig.5. The broadening described by $T_2^*$ results in a fast decay of the oscillation amplitude for short times $\tau_p$. This indicates that during the controlled rotations some of the spins will be dephased and will not participate in the collective motion, leading to a slight reduction of the Overhauser field, as it is being redirected. This explains the underestimated value of $T_{Rabi}$ when the curves in Fig.2 are fitted with a single exponential decay. Such spin loss can be overcome by increasing the power of rf excitation (i.e. $B_{rf}$), or by using dots with narrower nuclear spin resonances. The latter also prolongs the decay of the transverse component of the rotated Overhauser field following the full control pulse-sequence. Note, that such reduction of the effective coherence time $T_{Rabi}$ for single pulse rotations has not been observed in GaAs QWs \cite{Sanada}, as in those experiments $T_2^*=90 \mu$s was found, only 3 times shorter than $T_2=270 \mu$s indicating insignificant width of the NMR frequency distribution in GaAs quantum well samples. The broadening in the GaAs dots, which allow access to single electron spins, may occur due to local strain and hence quadrupole splittings due to the proximity to the GaAs/AlGaAs interface. 

To summarize, we report coherent manipulation of small ensembles of nuclear spins in optically pumped single GaAs/AlGaAs interface quantum dots using  sequences of radio frequency pulses. The reported full nuclear spin control in a dot can serve as a tool for fast manipulation of large effective magnetic fields up to a few hundred milli-Tesla and may be used for control of both electron and hole spin states on the microsecond time-scale. We also obtain further insight into the nuclear spin dynamics by studying spin dephasing. We find that transverse spin relaxation occurs on the time scale of $T_2^*=25\pm 2 \mu$s due to the inhomogeneous broadening of the resonance, but could be compensated by using spin-echo techniques leading to slower relaxation with an intrinsic coherence time $T_2=360\pm 40 \mu$s. We also note, that in optically pumped quantum dots where high nuclear spin polarizations of up to 60$\%$ are routinely achieved, the presented NMR techniques open the way to a new class of experiments in semiconductors, where optical and resonant rf techniques are be combined to generate highly-correlated nuclear spin states \cite{Slichter,Rudner}.\\

{\label{sec:Methods} \bf METHODS}

The sample investigated contains interface QDs formed naturally by 1 monolayer width fluctuations in a nominally 13 monolayer GaAs quantum well layer embedded in Al$_{0.33}$Ga$_{0.67}$As barriers (see growth details in Ref.\cite{Peter}). Relative to self-assembled dots, the near lattice-matched GaAs/AlGaAs QDs are highly favorable for NMR studies due to the weak strain and hence weak quadrupole interaction. We study neutral quantum dots. The ODNMR setup is sketched in Fig.1a \cite{Makhonin}. External magnetic field $B_z$ is applied in the Faraday geometry. Optical excitation is used (i) for pumping the nuclear spins in the dot via dynamic nuclear polarization \cite{Chekhovich1,Gammon1,Makhonin,Eble,Tartakovskii,Chekhovich2,Nikolaenko} and (ii) to excite photoluminescence (PL) for measurements of the exciton Zeeman splittings in individual dots. The samples were measured at a temperature $T=$4.2 K. We use an excitation laser at 670 nm which generates electrons and holes in the quantum well states $\approx$ 130 meV above the quantum dot emission lines. PL was detected with a double spectrometer and a charge coupled device. As shown in Fig.1a, a coil was wound around the sample for rf excitation of the dots. The coil was excited by the amplified output from an rf generator and provided transverse magnetic fields $B_{rf}$ up to $\approx$ 2 mT at the location of the dots. 

Fig.1b shows exciton PL spectra \cite{Nikolaenko} measured for $B_z = 2$T under $\sigma^+$ optical pumping. Pumping with circularly polarized light results in the dynamic nuclear polarization in the dot.  The resulting Overhauser field $B_N$ is detected through the change in the exciton Zeeman splitting, $\Delta E=g_e\mu_B B_N$ [$g_e$ electron g-factor, $\mu_B$ - Bohr magneton, $B_N$ is co-(anti-) parallel to $B_z$ for $\sigma^{-}$($\sigma^{+}$) excitation]. Using lineshape fitting $\Delta E$ is measured with an accuracy of $\approx$1$\mu$eV. rf excitation resonant with nuclear the spin transitions leads to transfer of population between the nuclear spin states and as a result to reduction of $|B_N|$. This is observed in Fig.1b as a change in the splitting of the Zeeman doublet when rf excitation is applied. \\

{\label{sec:Acknowledgments} \bf ACKNOWLEDGMENTS}

We thank L. M. K. Vandersypen and V. Fal'ko for fruitful discussions. This work has been supported by  the EPSRC Programme Grant EP/G601642/1 and the Royal Society.

\end{document}